----------
X-Sun-Data-Type: default
X-Sun-Data-Description: default
X-Sun-Data-Name: SSB
X-Sun-Charset: us-ascii
X-Sun-Content-Lines: 961

\magnification=1200
\settabs 18 \columns

\baselineskip=12 pt
\topinsert \vskip 1.25 in
\endinsert

\def\sqr#1#2{{\vcenter{\vbox{\hrule height.#2pt
 \hbox{\vrule width.#2pt height#1pt \kern#1pt
 \vrule width.#2pt} \hrule height.#2pt}}}}

\def\operp{\hbox{${\kern+.25em{\bigcirc}
\kern-.85em\bot\kern+.85em\kern-.25em}$}}

\def\lsim{\;\raise0.3ex\hbox{$<$\kern-0.75em\raise-1.1ex\hbox{$\sim$}}\;}
\def\gsim{\;\raise0.3ex\hbox{$>$\kern-0.75em\raise-1.1ex\hbox{$\sim$}}\;}
\def\no{\noindent}

\def\ce{\centerline}
\def\ve{\vfill\eject}
\def\rdots{\mathinner{\mkern1mu\raise1pt\vbox{\kern7pt\hbox{.}}\mkern2mu
 \raise4pt\hbox{.}\mkern2mu\raise7pt\hbox{.}\mkern1mu}}

\def\e e{$e^+ e^-$ }



\rightline{UCLA/99/TEP/30}
\rightline{June 1999}
\vskip1.0cm

\ce{{\bf SPONTANEOUS SYMMETRY BREAKING OF $q$-GAUGE FIELD THEORY}}
\vskip.5cm
\ce{\it R. J. Finkelstein}
\vskip.3cm
\ce{Department of Physics and Astronomy}
\ce{University of California, Los Angeles, CA  90095-1547}
\vskip1.0cm


\no {\bf Abstract.}  In non-Abelian field theories with $q$-symmetry groups
the massive 
particles have a non-local interpretation with a stringlike spectrum.
It is shown that a massless vector similarly acquires a tower of masses by
spontaneous symmetry breaking.  
\ve

\line{{\bf 1. Introduction.} \hfil}
\vskip.3cm

It is formally possible to generalize any non-Abelian field theory by replacing
its Lie group by the corresponding quantum group.  Since the Lie algebras are
completely rigid with respect to small deformations of the structure
constants, it is clear that any new theory obtained in this way must be very
different from the corresponding Lie theory and in particular it will contain
additional degrees of freedom.  While conventional Yang-Mills theories may
describe internal properties of point particles such as isotopic spin they
do not describe extension, i.e. solitonic or stringy properties of an 
elementary particle.  The quantum gauge theories may be able to do this$^1$ and
at the same time are approximated by a Yang-Mills theory in a correspondence
limit.  Here we pursue the correspondence further by considering spontaneous
symmetry breaking.
\vskip.5cm

\line{{\bf 2. The $q$-Yang-Mills Theory.} \hfil}
\vskip.3cm

The basic Lagrangian may be expressed as follows:$^1$
$$
S = \int d^4x\biggl\{-{1\over 4} Q_\ell F_{\mu\nu}F^{\mu\nu}Q_r +
i\psi^tC\epsilon\gamma^\mu\nabla_\mu\psi + {1\over 2}
\bigl[(\nabla_\mu\varphi)^t\epsilon(\nabla^\mu\varphi)+\varphi^t\epsilon
\varphi\bigr]\biggr\} \eqno(2.1)
$$
\no where kinetic terms in $Q_\ell$ and $Q_r$ have been omitted.  Here
$$
\eqalignno{Q_\ell^\prime &= Q_\ell T^{-1} \qquad
\psi^\prime = T\psi & (2.2) \cr
\noalign{\hbox{and}}
Q_r^\prime &= TQ_r \qquad ~~~\varphi^\prime = T\varphi~. & (2.3) \cr}
$$
\no $\psi$ is also a Dirac field while $\varphi$ is a scalar with respect to
the Lorentz group.  The covariant derivative transforms as
$$
\nabla_\mu^\prime = T\nabla_\mu T^{-1}~. \eqno(2.4)
$$
\no The gauge field $A_\mu$ is defined by
$$
A_\mu = \nabla_\mu-\partial_\mu \eqno(2.5)
$$
\no and transforms as
$$
A_\mu^\prime = TA_\mu T^{-1} + T\partial_\mu T^{-1}~. \eqno(2.6)
$$
\no The field strength is defined by
$$
F_{\mu\nu} = [\nabla_\mu,\nabla_\nu] \eqno(2.7)
$$
\no and transforms as
$$
F^\prime_{\mu\nu} = TF_{\mu\nu}T^{-1}~. \eqno(2.8)
$$
\no The fundamental new assumption is that the gauge group is $SL_q(2)$.
Then if $T$ belongs to $SL_q(2)$
$$
T^t\epsilon T = T\epsilon T^t = \epsilon \eqno(2.9)
$$
\no where $t$ means transpose and
$$
\epsilon = \left(\matrix{0 & q^{-1/2} \cr
-q^{1/2} & 0 \cr} \right)~. \eqno(2.10)
$$
\no In (2.1)
$$
Q = \left(\matrix{q^{-1} & 0 \cr 0 & q \cr} \right) \eqno(2.11)
$$
\no $\epsilon$ is analogous to the charge conjugation matrix $C$
intertwining the Lorentz transformations $L^t$ and $L^{-1}$
$$
L^t CL = LCL^t = C~. \eqno(2.12)
$$
\no For closure one requires
$$
(T_1T_2)^t \epsilon (T_1T_2) = T_2^t T_1^t\epsilon T_1T_2 = \epsilon
\eqno(2.13)
$$
\no which depends on
$$
(T_1T_2)^t = T_2^t T_1^t \eqno(2.14)
$$
\no but (2.14) will be satisfied in general only if the matrix elements of
$T_1$ commute with those of $T_2$.  To ensure this property one may take
$T_1$ and $T_2$ at spatially (causally) separated points with respect to the
light cone.  The resulting groupoid is therefore non-local.

We shall discuss the action (2.1) as the basis for a non-Abelian gauge
theory.  The treatment will differ from the conventional Lie theories since
the interacting fields in (2.1) are operator valued even before
quantization and it
is necessary to introduce not only the usual Fock space but also a second
state space.  To describe the theory more completely one must assign quantum
group representations to the constituent fields.  Finally in order to
compare with conventional theories we shall add a symmetry breaking term that
becomes important only near the minimum in the field energy.
\vskip.5cm

\line{{\bf 3. The Quantum Groups $SL_q(2),~SU_q(2)$, and $SO_q(2)$.} \hfil}
\vskip.3cm

Set
$$
T = \left(\matrix{a & b \cr c & d \cr} \right)~, \qquad
T\epsilon SL_q(2)~. \eqno(3.1)
$$
\no Then Eqs. (2.9) require that the matrix elements of $T$ satisfy the
following relations
$$
\eqalign{\hbox{(a)}~~ab &= qba \cr \hbox{(b)}~~ac &= qca \cr 
\hbox{(c)}~~bc &= cb \cr} \quad
\eqalign{\hbox{(d)}~~cd &= qdc \cr 
\hbox{(e)}~~bd &= qdb \cr \hfil \cr} \quad
\eqalign{
\hbox{(f)}~~&ad-qbc = 1 \cr \hbox{(g)}~~&ad-da=(q-q_1)bc \cr 
&q_1=q^{-1} \cr} \eqno(3.2)
$$
\no where $q$ may be complex.

A quantum subgroup is obtained by setting
$$
\eqalignno{d &= \bar a & (3.3a)\cr c &= -q_1\bar b~. & (3.3b) \cr} 
$$
\no Then
$$
\eqalign{\hbox{(a)}~~ab&=qba \cr \hbox{(b)}~~a\bar b&= q\bar ba \cr
\hbox{(c)}~~b\bar b &= \bar bb \cr} \quad
\eqalign{\hbox{(d)}~~\bar b\bar a &= q\bar a\bar b \cr
\hbox{(e)}~~b\bar a &= q\bar ab \cr \hfil \cr} \quad
\eqalign{\hbox{(f)}~~a\bar a &+ b\bar b=1 \cr
\hbox{(g)}~~\bar aa &+ q_1^2\bar bb = 1 \cr
\hfil \cr} \eqno(3.4) 
$$
\no Note that (3.4) is invariant under involution only if 
$$
\bar q = q~. \eqno(3.5)
$$
\no In the hermitian case, $\bar q=q^*$ (= complex conjugate) and by (3.5)
$q$ must be real.  The corresponding quantum group is $SU_q(2)$ or
$T^{-1} = (T^t)^*$. 

If, however, $\bar a=a^t,~\bar b=b^t$, and $\bar q=q^t$, where $t$ means
simple transposition, and if $q$ is a scalar matrix, then $\bar q=q^t=q$
where $q$ may be complex.  Then (3.4) and (3.5) are satisfied and the
corresponding quantum group is $SO_q(2)$ or $T^{-1} = T^t$.
\vskip.5cm

\line{{\bf 4. Representations of the $SU_q(2)$ and the $SO_q(2)$ Algebras.}
\hfil}
\vskip.3cm

A matrix representatiion of the algebra should be understood in the 
following discussion.
\vskip.3cm

\line{{(a) {\it Hermitian Involution $(SU_q(2))$}.} \hfil}
\vskip.3cm

\no Let
$$
T = \left(\matrix{\alpha & \beta \cr -q_1\bar\beta & \bar\alpha \cr} \right)
\eqno(4.1)
$$
\no Then permitting the operation of hermitian conjugation with real $q$ we may abbreviate (3.4) as follows:
$$
\eqalign{\alpha\beta &= q\beta\alpha \cr
\alpha\bar\beta &= q\bar\beta\alpha \cr
\beta\bar\beta &= \bar\beta\beta \cr} \qquad
\eqalign{&\alpha\bar\alpha + \beta\bar\beta = 1 \cr
&\bar\alpha\alpha + q_1^2\bar\beta\beta = 1 \cr
&q_1 = q^{-1} \cr} \eqno(4.2)
$$
\no If $q=1$, the equations (4.2) may be satisfied by complex numbers
$\alpha$ and $\beta$ subject to
$$
|\alpha|^2 + |\beta|^2 = 1~. \eqno(4.3)
$$
\no If $q=-1$, (4.2) may be satisfied by
$$
\eqalign{\bar\alpha &= \alpha = {1\over\sqrt{2}}
\left(\matrix{0 & 1 \cr 1 & 0 \cr} \right) = \sigma_1 \cr
\bar\beta &= \beta = {1\over\sqrt{2}}
\left(\matrix{0 & -i \cr i & 0 \cr}\right) = \sigma_2~. \cr} \eqno(4.4)
$$
\no In both cases, $q=\pm 1,~T\epsilon SU(2)$.
In addition to these finite representations there is an infinite
dimensional representation.  To derive this representation let us 
introduce the infinite dimensional state 
space associated with the algebra (4.2).

Denote the basis states by $|n\rangle,~n = 0,1,2~\ldots$.  Since $\beta$ and
$\bar\beta$ commute we may require
$$
\eqalignno{\alpha|0\rangle &= 0 & (4.5) \cr
\beta|0\rangle &= b|0\rangle & (4.6) \cr
\bar\beta|0\rangle &= b^*|0\rangle~, & (4.7) \cr}
$$
\no where $b$ amd $b^*$ are complex conjugates.
Then the remaining states are obtained with the aid of the raising
operator $\bar\alpha$ by
$$
\bar\alpha|n\rangle = \lambda_n|n+1\rangle~. \eqno(4.8)
$$ 
\no It is also consistent to set
$$
\alpha |n\rangle = \mu_n|n-1\rangle~. \eqno(4.9)
$$
\no Then the quadratic relations determine $\lambda_n$ and $\mu_n$ as
follows:
$$
\eqalignno{|\lambda_n| &= (1-|b|^2 q^{2n})^{{1\over 2}} & (4.10) \cr
\mu_n &= \lambda_{n-1} & (4.11) \cr}
$$
\no One also finds
$$
\eqalignno{\beta|n\rangle &= q^nb|n\rangle & (4.12) \cr
\bar\beta|n\rangle &= q^nb^*|n\rangle & (4.13) \cr
|b|^2 &= q^2 > 0 \to q~\hbox{is real}~. & (4.14) \cr}
$$
\no Eq. (4.10) implies negative norms unless $q^2\leq 1$.
The $|n\rangle$ are eigenstates of the self-adjoint operator
$$
H = {1\over 2} (\alpha\bar\alpha + \bar\alpha\alpha) \eqno(4.15)
$$
\no and are hence orthogonal.

The preceding operator representation may be displayed as an infinite matrix
representation in the usual way:
$$
\eqalignno{\langle n^\prime|\alpha|n\rangle &= \mu_n\delta(n^\prime,n-1)
& (4.16) \cr
\langle n^\prime|\bar\alpha|n\rangle &= \lambda_n\delta(n^\prime,n+1) 
& (4.17) \cr
\langle n^\prime|\beta|n\rangle &= q^nb\delta(n^\prime,n) & (4.18) \cr
\langle n^\prime|\bar\beta|n\rangle &= q^nb^*\delta(n^\prime,n) & (4.19) \cr}
$$
\no where $\lambda_n$ and $\mu_n$ are given by (4.10) and (4.11).
By restricting the action to a finite subspace of the state space one
may search for finite dimensional representations.  The 3-dimensional
representation illustrates the general case.
$$
\bar\alpha = \left(\matrix{0 & 0 & 0 \cr (1-q^2)^{1/2} & 0 & 0 \cr
0 & (1-q^4)^{1/2} & 0 \cr} \right) \qquad
\alpha = \left(\matrix{0 & (1-q^2)^{1/2} & 0 \cr
0 & 0 & (1-q^4)^{1/2} \cr 0 & 0 & 0 \cr} \right) \eqno(4.20)
$$
$$
\beta = \left(\matrix{q & 0 & 0 \cr 0 & q^2 & 0 \cr
0 & 0 & q^3 \cr} \right) \qquad \bar\beta = \beta^*~. \eqno(4.21)
$$
\no Then $\alpha\bar\alpha + \beta\bar\beta = 1$ implies
$$
\eqalign{
&\left(\matrix{1 & 0 & 0 \cr 0 & 1 & 0 \cr 0 & 0 & q^6 \cr} \right) =
\left(\matrix{1 & 0 & 0 \cr 0 & 1 & 0 \cr 0 & 0 & 1 \cr} \right) \cr
&\hbox{or}~q^6 = 1~ 
\hbox{and}~q = \exp\biggl[\biggl({2\pi i\over 6}\biggr)n\biggr] \cr}
\eqno(4.22)
$$
\no But $q$ must be real, therefore
$$
q=\pm 1~. \eqno(4.23)
$$

There are no additional finite-dimensional representations to be obtained
in this way.
\vskip.5cm

\line{{(b) {\it Transposition Involution $(SO_q(2))$.} } \hfil}
\vskip.3cm

The matrix equations (3.3) and (3.4) and (4.2) may be read 
slightly differently by
substituting simple transposition for hermitian conjugation.  Then the
equations (4.2) retain their same form but now
$$
\eqalign{\bar\alpha &= \alpha^t \cr
\bar\beta &= \beta^t = \beta \cr \bar q &= q^t = q \cr} \eqno(4.24)
$$
\no and $q$ may be complex.

The equations describing the state space are also the same except for changes
required by (4.24).  For example
$$
\eqalign{\beta\bar\beta|0\rangle &= b^2|0\rangle \cr
b^2 &= q^2 \cr} \quad
\eqalign{&\hbox{instead of} \cr
&\hbox{instead of} \cr} \quad
\eqalign{\beta\bar\beta|0\rangle &= |b|^2|0\rangle \cr
|b|^2 &= q^2 \cr} \eqno(4.25)
$$

The principal equations now read as follows:
$$
\eqalign{\alpha|0\rangle &= 0 \cr \beta|0\rangle &= b|0\rangle \cr
\bar\beta|0\rangle &= b|0\rangle \cr
\bar\alpha|n\rangle &= \lambda_n|n+1\rangle \cr
\alpha|n\rangle &= \mu_n|n-1\rangle~, \cr} \eqno(4.26)
$$
\no where
$$
\eqalign{&\lambda_n = (1-q^{2n+2})^{{1\over 2}}, \cr
&\mu_n = \lambda_{n-1}~. \cr}
$$
$$
\eqalign{\langle n^\prime|\alpha|n\rangle &=
\mu_n\delta(n^\prime,n-1) \cr
\langle n^\prime|\bar\alpha|n\rangle &= \lambda_n\delta(n^\prime,n+1) \cr
\langle n^\prime|\beta|n\rangle &= q^{n+1}\delta(n^\prime,n) \cr}
\eqno(4.27)
$$
\no Eqs. (4.20) and (4.21) for the 3-dimensional representation are also
unchanged except that $\bar\beta = \beta^t$ and $q$ may now be complex.

Since the complex roots of unity are now allowed, there is now an infinite
class of finite-dimensional representations.  In the $N$-dimensional
representation
$$
\beta = \left(\matrix{q & \hfil & \hfil & \hfil \cr
\hfil & q^2 & \hfil & \hfil \cr 
\hfil & \hfil & \ddots & \hfil \cr
\hfil & \hfil & \hfil & q^N \cr} \right)~,
\qquad q = \exp\biggl[{\pi i\over N}\biggr]~. \eqno(4.28)
$$
\no For example, if $N=2$,
$$
q=\exp\biggl[{\pi i\over 2}\biggr] = i \eqno(4.29)
$$
$$
\alpha = \left(\matrix{0 & \sqrt{2} \cr 0 & 0 \cr} \right) \quad
\bar\alpha = \left(\matrix{0 & 0 \cr \sqrt{2} & 0 \cr} \right) \quad
\beta = \left(\matrix{i & 0 \cr 0 & i^2 \cr} \right)~. \eqno(4.30)
$$

There is the following correspondence 
between the $q$-algebra and the
conventional notation for the Cartan subalgebra in both cases (a) and (b):
$$
\eqalign{\bar\alpha &\sim E_+ \cr
\alpha &\sim E_- \cr
\beta,\bar\beta &\sim H~. \cr}  \eqno(4.31)
$$
\vskip.5cm

\line{{\bf 5. A Different Reduction of the State Space of $SL_q(2)$.} \hfil}
\vskip.3cm

By not making the second assumption (3.3b) one obtains a less restrictive
state space.

Eqs. (3.2) may be rewritten to bring out the following algebraic 
automorphism: \break $\alpha\leftrightarrow\delta,~q\leftrightarrow q_1$.
$$
\eqalign{\alpha\beta &= q\beta\alpha \cr
\delta\beta &= q_1\beta\delta \cr} \quad
\eqalign{\alpha\gamma &= q\gamma\alpha \cr
\delta\gamma &= q_1\gamma\delta \cr} \quad
\eqalign{\alpha\delta &- q\beta\gamma = 1 \cr
\delta\alpha &- q_1\beta\gamma = 1 \cr} \quad
\eqalign{\beta\gamma &= \gamma\beta \cr
\hfil \cr} \eqno(5.1) 
$$
\no Since $\beta$ and $\gamma$ commute they have a common set of eigenstates.
Let the ground state, $|0\rangle$, be defined by
$$
\eqalign{\alpha|0\rangle &= 0\cr
\beta|0\rangle &= b_o|0\rangle \cr
\gamma|0\rangle &= c_o|0\rangle~. \cr} \eqno(5.2)
$$
\no Define the state $|n\rangle$ by the recursive relations:
$$
\eqalign{\delta|n\rangle &= \lambda_n|n+1\rangle \cr
\alpha|n\rangle &= \mu_n|n-1\rangle \cr} \eqno(5.3)
$$
\no Then by (5.1)
$$
\eqalign{\beta|n\rangle &= q^nb_o|n\rangle \cr
\gamma|n\rangle &= q^nc_o|n\rangle~. \cr} \eqno(5.4)
$$
\no The algebraic symmetry may be used to define the dual states according
to
$$
\eqalign{\langle 0|\delta &= 0 \cr
\langle n|\alpha &= \langle n+1|\lambda_n \cr
\langle n|\delta &= \langle n-1|\mu_n~. \cr} \eqno(5.5)
$$
\no We also assume
$$
\eqalign{\langle 0|\beta &= \langle 0|b_o \cr
\langle 0|\gamma &= \langle 0|c_o \cr} \eqno(5.6)
$$
\no Then by (4.1)
$$
\eqalign{\langle n|\beta &= \langle n|q^nb_o \cr
\langle n|\gamma &= \langle n|q^nc_o \cr} \eqno(5.7)
$$
\no Set
$$
\eqalign{\delta &= \bar\alpha \cr
\beta &= \bar\beta \cr
\gamma &= \bar\gamma \cr} \eqno(5.8)
$$
\no where the bar means either (a) hermitian 
or (b) transposition conjugation.  

We discuss (a) first.  Then the dual equations
(5.5)-(5.7) are also hermitian conjugates of (5.2)-(5.4) if $q$ and the other
numerical factors are all real as we shall assume.

We also have by (5.1)
$$
\eqalignno{\bar\alpha\alpha -q_1\beta\gamma &= 1 & (5.9a) \cr
\alpha\bar\alpha -q\beta\gamma &= 1 & (5.9b) \cr}
$$
\no Under the algebraic automorphism (5.9a) $\leftrightarrow$ (5.9b) and under
hermitian conjugation, these equations go into themselves.

We may compute
$$
\langle n|\alpha\bar\alpha|n\rangle = \langle n|\alpha\lambda_n|n+1\rangle =
\lambda_n\mu_{n+1}\langle n|n\rangle = \lambda_n\mu_{n+1} \eqno(5.10)
$$
\no and also by (5.3), (5.5) and (5.8)
$$
\langle n|\alpha\bar\alpha|n\rangle =
\langle n|\alpha\cdot\bar\alpha|n\rangle =
|\lambda_n|^2\langle n+1|n+1\rangle = |\lambda_n|^2~. \eqno(5.11)
$$
\no where we have set $\langle n|n\rangle = 1$ for all $n$.
Then
$$
\lambda_n^* = \mu_{n+1}~. \eqno(5.12)
$$
\no By (5.9a)
$$
b_oc_o = q~. \eqno(5.13)
$$
\no By (5.9b)
$$
\langle n|\alpha\bar\alpha - q\beta\gamma|n\rangle = 1
$$
\no and by (5.11) and (5.13)
$$
|\lambda_n|^2 - q~q^{2n}b_oc_o = 1~.
$$
\no Then
$$
\eqalignno{|\lambda_n|^2 &= 1+q^{2n+2} & (5.14) \cr
|\mu_n|^2 &= 1+q^{2n} & (5.15) \cr}
$$
\no by (5.12).

In case (b) the bar in (5.8) and following equations means transposition 
conjugation.  The dual state equations are also transposition conjugate.
Then $q$ and other numerical factors are not constrained to be real.
The scalar products, measured by $\lambda_n^2$ rather than $|\lambda_n|^2$,
may still be chosen positive for a proper choice of $q$:
$$
\lambda_n^2 = 1+q^{2n+1}~. \eqno(5.16)
$$

In reductions of $SL_q(2)$ to $SU_q(2)$ and $SO_q(2)$ we have assumed
(3.3b).  In that case one must also assume $q^2\leq 1$ according to
(4.10) and (4.27), but we are now assuming that $\gamma$ and $\beta$ are
separately hermitian (symmetric) 
rather than hermitian (symmetric) conjugates and $q$ may therefore
be greater than unity by (5.14) and (5.16).

In terms of $T$ the equation (5.8) means
$$
\bar T \sigma_1T^{-1} = \sigma_1 \eqno(5.17)
$$
\no where
$$
\sigma_1 = \left(\matrix{ 0 & 1 \cr 1 & 0 \cr} \right) \quad \hbox{and} \quad
T = \left(\matrix{\alpha & \beta \cr \gamma & \delta \cr} \right)~.
\eqno(5.18)
$$

\vskip.5cm

\line{{\bf 6. The Choice of Gauge.} \hfil}
\vskip.3cm

Since the theory is $T$-invariant, all fields may lie in the algebra (3.2).
Every $T$-transformation will introduce new powers of the generators
$(\alpha,\bar\alpha,\beta,\bar\beta)$.  The theory may be completed by
introducing a privileged gauge that breaks the $T$-invariance.  In the
Higgs model the symmetry breaking is accomplished by associating the
privileged gauge with the minimum in the field energy.  One may imagine
a similar mechanism here.  In any case we would like the $q$-theory to lie
close enough to the Lie theory to permit a plausible correspondence limit.
In the Lie case $(SU(2))$ the vector and neutral scalar fields may be
written
$$
\eqalignno{W_\mu &= W_\mu^+\tau_- + W_\mu^-\tau_+ + W_\mu^3\tau_3 
 & (6.1) \cr
\Phi &= \Phi^3\tau_3 + \Phi^0 1 ~. & (6.2) \cr}
$$

To go over from the Lie algebra to the $q$-algebra we adopt the
correspondence (4.31) and write
$$
\eqalignno{W_\mu &= W_\mu^+\alpha + W_\mu^-\bar\alpha +
W_\mu^\beta\beta + W_\mu^{\bar\beta}\bar\beta & (6.3) \cr
\Phi &= \Phi^\beta\beta + \Phi^{\bar\beta}\bar\beta & (6.4) \cr}
$$
\no since the diagonal elements correspond to $(\beta,\bar\beta)$.

In the $SU_q(2)$ theory $q$ is real $q^2\leq 1$, and
$(\alpha,\bar\alpha,\beta,\bar\beta)$ are regarded as operators on the
infinite dimensional internal state space.  
Then there are 4 kinds of $W$ particles:
$W^+,W^-,W^\beta$, and $W^{\bar\beta}$, and there are 2 kinds of $\Phi$
particles: $\Phi^\beta$ and $\Phi^{\bar\beta}$; and each of these carries
an infinite set of excited states labelled by the eigenvalue of $\beta$.

In the $SO_q(2)$ theory the transition from $SL_q(2)$ to a subgroup is
accomplished with the aid of the transposition involution (which is the more
economical choice since $SL_q(2)$ is itself defined in terms of a transposition).  In this case, $q$ is an $N^{\rm th}$ root of unity.  Then
$\beta = \bar\beta$ and each of the 4 particles $(W^+,W^-,W^\beta,\Phi^\beta)$
carries a finite set of excited states (of height $N$ since the
eigenvalues of $\beta$ will run from $q$ to $q^N$).

One next has the option of regarding $q$ (a) as a basic parameter of the
second theory or (b) as a new quantum number that will be different for
the different particle multiplets.  With option (b) one would be able to
include the $q=\pm 1$ particles, the familiar $SU(2)$ particles.
In this case
we may regard $q$ as an operator commuting with the rest of the
algebra and equal to the 
product of a hermitian and a unitary
matrix so that its eigenvalues may be either real or roots of unity, 
depending on the state on which it operates.  In this way one would arrive
at a more general formalism that would subsume the separate cases already
mentioned.  The enlarged state space is then divided into disjoint sectors
labelled by the eigenvalues of $q$.  Since $q$ commutes with the complete
algebra, transitions between the separate sectors are not possible within
the algebra.

In the $SL_q(2)$ case $\beta$ and $\gamma$ are separately hermitian rather
than hermitian conjugates.  There are again 4 kinds of vector particles
but there are no finite representations.  The important distinction is now
that $q^2\leq 1$ no longer holds.
\vskip.5cm

\line{{\bf 7. Vector Masses for $SU_q(2)$ and $SO_q(2)$.} \hfil}
\vskip.3cm

Following the usual ideas about spontaneous symmetry breaking, we may assume
a neutral scalar field, a functional potential leading to a non-invariant
lowest state, the existence of Goldstone bosons if the symmetry is global,
and the replacement of Goldstone bosons by massive vectors if the symmetry
is position-dependent.  We shall accordingly add to the original Lagrangian
a symmetry breaking term that becomes important near the minimum in the
field energy and couples a neutral scalar field to the vector field.

The interaction term is assumed to be invariant under $SU(2)$ only,
while the interacting fields continue to lie in the $SU_q(2)$ or the $SO_q(2)$
algebra.  Then the vector mass term is buried in the following kinetic
energy term:
$$
(\nabla_\mu\Phi)^\dagger (\nabla^\mu\Phi) \eqno(7.1)
$$
\no where the dagger indicates hermitian conjugation.  Here $\nabla_\mu$
is the covariant derivative:
$$
\nabla_\mu = \partial_\mu + W_\mu(+)\alpha + W_\mu(-)\bar\alpha +
W^\beta_\mu\beta + W_\mu^{\bar\beta}\bar\beta \eqno(7.2)
$$
\no and $\Phi$ is a neutral scalar field:
$$
\Phi = (\mathrel{\mathop\varphi^{\circ}} + 
\mathrel{\mathop\varphi^{\scriptstyle 1}} + \ldots)\beta \eqno(7.3)
$$
\no where $\displaystyle\mathrel{\mathop\varphi^{\circ}}$
minimizes the energy and determines the vacuum expectation value.
$\displaystyle\mathrel{\mathop\varphi^{\circ}}$ will be taken real.

We are interested only in the lowest order contribution to (7.1) and 
only in terms determining the vector masses.  Then the terms contributing
to the mass are
$$
\nabla_\mu\Phi \to (W_\mu(+)\alpha\beta + W_\mu(-)
\bar\alpha\beta + W_\mu^\beta\beta^2)\mathrel{\mathop\varphi^{\circ}}
\eqno(7.4)
$$
\no where for simplicity the last term in (7.2) has also been dropped
and
$$
(\nabla_\mu\Phi)^\dagger \to (W_\mu(-)\beta^+\alpha^+ +
W_\mu(+)\beta^+\bar\alpha^+ + (W_\mu^\beta)^*(\beta^2)^+)
\mathrel{\mathop\varphi^{\circ}}~. \eqno(7.5)
$$
\no To keep a correspondence with the standard case we have set
$$
W_\mu(\pm) = {1\over 2} (W_{1\mu} \pm i~W_{2\mu})~. \eqno(7.6)
$$
\no The expectation value of (7.1) in the $n^{\rm th}$ state contributes
$$
\langle n|W_\mu(-)W^\mu(+)\beta^+\alpha^+\alpha\beta +
W_\mu(+) W^\mu(-)\beta^+\bar\alpha^+\bar\alpha\beta + (W_\mu^\beta)^*
(W^\beta)^\mu(\beta^2)^+\beta^2|n\rangle \varphi_o^2~. \eqno(7.7)
$$
\no There are two cases:
\vskip.3cm
\item{(a)} hermitian conjugation~~~~, ~~$q$ is real and the group is $SU_q(2)$
\item{(b)} transposition conjugation, ~~$q$ is root of unity and the group is
$SO_q(2)$~.
\vskip.3cm

\no In case (a) expression (7.7) becomes
$$
\langle n|W_\mu(-)W^\mu(+)\bar\beta\bar\alpha\alpha\beta +
W_\mu(+)W^\mu(-)\bar\beta\alpha\bar\alpha\beta + (W_\mu^\beta)^*
(W^\beta)^\mu(\beta\bar\beta)^2|n\rangle\varphi_o^2~. \eqno(7.8)
$$
\no But
$$
W_\mu(-) W^\mu(+) = {1\over 4} (W_1^2 + W_2^2)~. \eqno(7.9)
$$
\no Write
$$
W_\mu^\beta = \hbox{\bf Z}_\mu~. \eqno(7.10)
$$
\no Then (7.8) becomes
$$
\biggl[{1\over 4}(W_1^2 + W_2^2)\langle n|\bar\beta(\bar\alpha\alpha + 
\alpha\bar\alpha)\beta|n\rangle + 
\hbox{\bf Z}^2\langle n|(\bar\beta\beta)^2|n\rangle\biggr]\varphi_o^2
\eqno(7.11)
$$
\no where
$$
\langle n|\bar\beta(\bar\alpha \alpha + \alpha\bar\alpha)\beta|n\rangle =
\langle n|2\bar\beta\beta - (1+q_1^2)(\bar\beta\beta)^2|n\rangle \eqno(7.12)
$$
\no and
$$
\langle n|\bar\beta\beta|n\rangle = q^{2n+2}~. \eqno(7.13)
$$
\no Therefore (7.1) is reduced to
$$
M^2_W{1\over 4}(W_1^2 + W_2^2) + M_z\hbox{\bf Z}^2 \eqno(7.14)
$$
\no where
$$
M_W^2 = {1\over 4}(2q^{2n+2}-q^{4n+4}-q^{4n+2})
\varphi_o^2 \eqno(7.15)
$$
\no and
$$
M^2_{\bf Z} = q^{4n+4}\varphi_o^2~. \eqno(7.16)
$$
\no Here $M_W$ and $M_{\bf Z}$ are the masses of the three vectors in
this model.

In case (b) expression (7.7) becomes
$$
\langle n|W_\mu(-)W^\mu(+)\beta^\dagger\alpha^\dagger\alpha\beta +
W_\mu(+)W^\mu(-)\beta^\dagger\alpha^*\bar\alpha\beta +
(W_\mu^\beta)^*(W^\beta)^\mu(\beta^\dagger\beta)^2|n\rangle
\varphi_o^2~. \eqno(7.17)
$$
\no To reduce (7.17) we need $\langle n|\beta^\dagger\alpha^\dagger\alpha\beta|n\rangle,~\langle n|\beta^\dagger\alpha^*\bar\alpha\beta|n\rangle$ and 
$\langle n|(\beta^\dagger\beta)^2|n\rangle$.  Since the eigenvalues of
$\beta$ are the roots of unity, one has
$$
\beta^\dagger\beta = 1~. \eqno(7.18)
$$
\no Then
$$
\eqalignno{\beta^\dagger\alpha^\dagger\alpha\beta &=
\alpha^\dagger\beta^\dagger\beta\alpha = \alpha^\dagger\alpha & (7.19) \cr
\beta^\dagger\alpha^*\bar\alpha\beta &= \alpha^*(\beta^\dagger\beta)\bar\alpha
= \alpha^*\bar\alpha~. & (7.20) \cr}
$$
\no By (4.27)
$$
\eqalign{\langle n|\alpha^\dagger\alpha|n\rangle &=
\sum_m\langle n|\bar\alpha|m\rangle^*\langle m|\alpha|n\rangle \cr
&= |\lambda_{n-1}|^2 \cr
\langle n|\alpha^*\alpha^t|n\rangle &= |\lambda_n|^2 \cr} \eqno(7.21)
$$
\no and
$$
\eqalign{|\lambda_n|^2 &= \bigl[(1-q^{2n+2})(1-q_1^{2n+2})\bigr]^{{1\over 2}} \quad (q_1 = q^{-1} = q^*) \cr
&= \bigl[2-(q^{2n+2} + q_1^{2n+2})\bigr]^{{1\over 2}}~. \cr} \eqno(7.22)
$$
\no By (4.28)
$$
|\lambda_n|^2 = 2\biggl|\sin{\pi\over N} (n+1)\biggr|~. \eqno(7.23)
$$
\no The expression (7.17) now becomes
$$
\biggl[{1\over 4}(W_1^2 + W_2^2)(|\lambda_n|^2 + |\lambda_{n-1}|^2) +
\hbox{\bf Z}^2\biggr]\varphi^2_o \eqno(7.24)
$$
\no so that the masses appearing in the $N$-dimensional multiplet
are
$$
\eqalignno{&M_W(n)^2 = {1\over 2}\varphi_o^2
\biggl[\biggl|\biggl(\sin{\pi\over N}(n+1)\biggr)\biggr| +
\biggl|\biggl(\sin{\pi\over N} n\biggr)\biggr|\biggr] & (7.25) \cr
&M_{\bf Z}^2 = \varphi_o^2~. & (7.26) \cr}
$$

\ve

\line{{\bf 8. Vector Masses: $SL_q(2)$ with $\bar T\sigma_1=\sigma_1T$.} \hfil}
\vskip.3cm

We discuss only case (a).  Then 
$\beta$ and $\gamma$ are separately hermitian rather than
hermitian conjugates and $q$ is real but otherwise unrestricted.  We repeat
the calculation of mass for this case.

Dropping the vector index one has
$$
\nabla = \partial + C + N \eqno(8.1)
$$
$$
\eqalign{C &= W(+)\alpha + W(-)\bar\alpha \cr
N &= W_\beta\beta + W_\gamma\gamma \cr
\Phi &= \rho_\beta\beta + \rho_\gamma\gamma~. \cr} \eqno(8.2)
$$
\no The mass terms are
$$
\langle n|\bigl[(C+N)\Phi\bigr]^\dagger\bigl[(C+N)\Phi\bigr]|n\rangle =
\langle n|\Phi^\dagger (C^\dagger C+N^\dagger N)\Phi|n\rangle \eqno(8.3)
$$
\no where
$$
C^\dagger C = {1\over 4}(W_1^2+W_2^2)[2+(q+q_1)\beta\gamma]~. \eqno(8.4)
$$
\no Since $C^+C$ amd $N^+N$ depend only on $\beta$ and $\gamma$ they commute 
with $\Phi$ and $\Phi^\dagger+$.  Then (8.3) becomes
$$
\langle n|(C^\dagger C+N^\dagger N)\Phi^\dagger\Phi|n\rangle \eqno(8.5)
$$
\no and the contribution of the ``charged" vectors is
$$
\langle n|C^\dagger C|n\rangle\langle n|\Phi^\dagger\Phi|n\rangle =
{1\over 4}(W_1^2+W_2^2)[2+(q+q_1)q^{2n+1}]
\langle n|\Phi^\dagger\Phi|n\rangle~. \eqno(8.6)
$$
\no Hence the mass$^2$ of the ``charged" vectors is
$$
M_n(\pm)^2 = {1\over 4}[2+q^{2n+2}+q^{2n}]\langle n|\Phi^\dagger\Phi|n\rangle~.
\eqno(8.7)
$$
\no The contribution of the neutral fields is
$$
\eqalign{\langle n|N^\dagger N|n\rangle\langle n|\Phi^\dagger\Phi|n\rangle &=
(W_\beta b_o+W_\gamma c_o)^2 q^{2n}
\langle n|\Phi^\dagger\Phi|n\rangle \cr
&= {\bf Z}^2 q^{2n}\langle n|\Phi^\dagger\Phi|n\rangle \cr} \eqno(8.8)
$$
\no where
$$
{\bf Z} = W_\beta b_o + W_\gamma c_o \eqno(8.9)
$$
\no and the corresponding mass$^2$ is
$$
M^2_n({\bf Z}) = q^{2n}\langle n|\Phi^\dagger\Phi|n\rangle~. \eqno(8.10)
$$
\no The ratio of the masses of the charged and neutral vectors is then
$$
\eqalign{\biggl({M_\pm\over M_{\bf Z}}\biggr)_n &= {1\over 2}
(2q^{-2n} + q^2+1) \cr
\biggl({M_\pm\over M_{\bf Z}}\biggr)_o &= {1\over 2}(3+q^2)~. \cr} \eqno(8.11)
$$
\no By (8.2) $N$ contains a $\beta$-field and also a $\gamma$-field; but
since the eigenvalues of $\beta$ and $\gamma$ are proportional by (5.4),
one of these fields is redundant and permits the replacement of $W_\beta$
and $W_\gamma$ by {\bf Z} in Eq. (8.9).

There is always enough information in the lowest lying multiplet to fix $q$.
If $q$ should turn out to be real and exceed unity in the case of other higher
vector multiplets, then the formalism would suggest a phenomenological theory
predicting the manner in which the low-lying pattern would be replicated at
higher energies.
\vskip.5cm

\line{{\bf 9. Summary.} \hfil}
\vskip.3cm

We have discussed a formal extension of non-Abelian field theory with the
following properties: the symmetry group is non-local; the dynamical fields
do not lie in a Lie algebra.  As a consequence, when the fields are resolved
in normal modes, it is necessary to expand state space in order to make a
particle interpretation.  It has been shown earlier$^1$ for free scalar fields
that the particles associated with the expanded state space are characterized
by a string-like spectrum.

If $q$ is real, the $SL_q(2)$ group is collapsed to the $SU_q(2)$ group;
if $q$ is a root of unity $SL_q(2)$ is collapsed to $SO_q(2)$.  If $q$ is
a root of unity, the particle multiplets are finite-dimensional; but if $q$
is real, only an infinite-dimensional multiplet is allowed.  In order to
accommodate both kinds of multiplets, it is necessary that total state
space be divided into disjoint sectors corresponding to different
values of $q$.  Transition btween sectors can be mediated by additional
interactions lying outside the $q$-algebra.  In both the $SU_q(2)$ and
$SO_q(2)$ examples $q^2\leq 1$.  To obtain the possibility of $q>1$ and real
one may utilize $SL_q(2)$ with $\bar T\sigma_1=\sigma_1T$.

The principal novelty of this approach lies in the alternative that it provides
for introducing into field theory non-locality and solitonic particles in
an algebraic rather than a geometric framework such as proposed in
stringlike or other higher dimensional theories.
\vskip.5cm

\line{{\bf Acknowledgements.} \hfil}
\vskip.3cm

I thank Professors Fronsdal and Varadarajan for comments.
\vskip.5cm

\line{{\bf References.} \hfil}
\vskip.3cm

\item{1,} R. Finkelstein, ``Gauged $q$-Fields", hep-th/9906135;
``Observable Properties of $q$-Deformed Physical Systems", hep-th/9906136.

\bye
\end